\documentclass[aps,nofootinbib]{revtex4}    % PRD
\usepackage{graphics}
\usepackage{epsfig}
\usepackage{array}
\def\apj{{\sl Astrophys.\ J. \ }}

\def\apjs{{\sl Astrophys.\ J.\ Supp. \ }}

\def\astroph#1{{\tt astro-ph/#1}}
\def\cqg{{Class.\ Quant.\ Grav. \ }}

\def\n{{\sl Nature \ }}

\def\np{{\sl Nucl.\ Phys. \ }}

\def\prd{{\sl Phys.\ Rev.\ D \ }}

\def\prl{{\sl Phys.\ Rev.\ Lett. \ }}

\def\ptps{{Prog.\ Theor.\ Phys.\ Suppl. \ }}

\newcommand{\gsim}{\,\lower2truept\hbox{${>\atop\hbox{\raise4truept\hbox{$\sim$}}}$}\,}
\newcommand{\pp}{~~~.}
\newcommand{\vv}{~~~,}

\newcommand{\be}{\begin{equation}}
\newcommand{\ee}{\end{equation}}
\newcommand{\bea}{\begin{eqnarray}}
\newcommand{\eea}{\end{eqnarray}}

\begin{document}

\title[Coupled and Extended Quintessence]{Coupled and Extended Quintessence: theoretical differences and structure formation}

\author{Valeria Pettorino$^{1}$, Carlo Baccigalupi$^{2}$
\\ $^{1}$ Institut f\"{u}r Theoretische Physik, Universit\"{a}t Heidelberg,
   Philosophenweg 16, D-69120 Heidelberg, Germany.
\\ $^{2}$ SISSA/ISAS, Via Beirut 4, I-34014 Trieste, and
\\ INFN, Sezione di Trieste, Via Valerio 2, I-34127 Trieste, Italy.
}
\begin{abstract} The case of a coupling between dark energy and matter (Coupled
Quintessence) or gravity
(Extended Quintessence) has recently attracted a deep interest and has
been widely investigated
both in the Einstein and in the Jordan frames (EF, JF), within scalar
tensor theories. Focusing
on the simplest models proposed so far, in this paper we study the
relation existing between the
two scenarios, isolating the Weyl scaling which allows to express them in
the EF and JF. 
Moreover, we perform a
comparative study of the behavior
of linear perturbations in both scenarios, which turn out to behave in a
markedly different way.
In particular, while the clustering is enhanced in the considered CQ models with 
respect to the corresponding Quintessence ones where the coupling is absent and 
to the ordinary cosmologies with a Cosmological Constant and Cold Dark Matter
($\Lambda$CDM), structures in EQ models may grow slower. 
This is likely to have direct consequences on the inner properties of 
non-linear structures, like cluster concentration, as well as on the weak lensing 
shear on large scales. Finally, we specialize our study for interfacing linear 
dynamics and N-body simulations in these cosmologies, giving a recipe for the 
corrections to be included in N-body codes in order to take into account the 
modifications to the expansion rate, growth of structures, and strength of gravity. 
\end{abstract}

\maketitle

\section{Introduction}
\label{i} 

The increasing amount of cosmological observations has been recently
pointing out that the Universe is
nearly spatially flat, with an expansion rate of about 70 km/s/Mpc and
with a $76 \%$ contribution to the
total energy density due to a non-clustered, negative pressure component
responsible for the acceleration
in the cosmic expansion, known as dark energy. The energy level of this
component is markedly lower than
almost all known physical quantities, except perhaps the neutrino mass
differences, and this motivated
big efforts in order to understand its physical properties. Interesting
hints have been obtained studying
scenarios in which the dark energy is dynamic, i.e. described by an
evolving and fluctuating scalar field,
the Quintessence. In particular, several investigations have been devoted
to study the possibility that the
dark energy might be coupled to other entities in the Universe, altering
the dynamics of the Quintessence field
itself as well as the evolution of structure formation. It is then
essential to understand the impact of such
a coupling on both cosmological expansion and perturbation dynamics, in
view of a clear prediction of observable
effects for the next to come experiments.
\\
Two main efforts have been recently carried out in a nearly parallel way.
On one side, within the usual
framework of General Relativity (Einstein Frame, EF) the effect of a
coupling between dark energy and dark matter
(Coupled Quintessence, CQ) has been deeply investigated both in the linear
\cite{amendola_2000} and non-linear regimes \cite{amendola_2004, matarrese_etal_2003},
 including the effects
of such models on the Cosmic Microwave Background (CMB)
\cite{lee_etal_2006, wang_etal_2007, mainini_etal_2007_2}, on Supernovae
\cite{amendola_etal_2004, amendola_etal_2006, guo_etal_2007}, on the matter power
spectrum \cite{mainini_etal_2007}, on structure
formation \cite{abdalla_etal_2007, bertolami_etal_2007}, also via N-body
simulation \cite{maccio_etal_2004}, as well as different choices of the coupling 
\cite{brookfield_etal_2007, mangano_etal_2003, koivisto_2005, quartin_etal_2008, olivares_etal_2008, boehmer_etal_2008}.
The possibility of an interaction between dark energy and neutrinos has also
been investigated \cite{fardon_etal_2004, brookfield_etal_2006},
leading recently to the proposal of growing matter scenarios
\cite{amendola_etal_2007, wetterich_2007, mota_etal_2008}.
Instabilities within coupled dark energy theories have also been addressed
\cite{bean_etal_2007, afshordi_2005, bjaelde_2007}.\\
On the other side, a wide variety of analysis has been carried out for
evaluating the possibility that the gravity
perceived by dark energy and matter deviates from General Relativity (GR). In
this latter case, extensions of GR have been
considered, in which the dark energy might derive from a non-minimal
interaction to gravity via an explicit coupling
between Quintessence and the Ricci scalar (Jordan Frame, JF). This is the
case of scalar tensor theories
\cite{hwang_1990, hwang_1990_2, faraoni_2000, schimd_etal_2005, riazuelo_uzan_2002, uzan_1999,
perrotta_etal_2000, boisseau_etal_2000, perrotta_baccigalupi_2002, matarrese_etal_2004,
pettorino_etal_2005, pettorino_etal_2004, perrotta_etal_2004},
which are known as Extended Quintessence (EQ) scenarios in the framework
of dark energy.
One of the most interesting results in this framework consists in the so
called `gravitational dragging'
\cite{perrotta_baccigalupi_2002}: the coupling to gravity introduces an
effective potential in which the
quintessence scalar field rolls, acquiring a dynamics which is a purely
gravitational effect. Eventually,
EQ energy density perturbations can be dragged to cluster through the
coupling with the
gravitational potential activated by the dark matter, its density
perturbations reaching non-linearity
at scales and redshifts relevant for the structure formation process. 
In particular, the growth of dark energy perturbations in the non-linear regime
 within EQ has been addressed in \cite{perrotta_etal_2004}. \\
Furthermore, in the case in which, in the EF, dark energy couples
universally to all matter fields,
the two parallel investigations can also be conformally related by Weyl
scaling, which allows to go
univoquely from one representation to the other. Although the resulting
physical effects must be equivalently
present in both frames, some features might more naturally appear in one
representation than in the other and
viceversa.
\\
In this paper, we identify the theoretical relation between the simplest
and most popular examples of both
coupling choices for CQ and EQ. On the basis of this analysis, we work out
the main phenomenological differences
in terms of the growth of linear perturbations, which turn out to behave
in a markedly different way for the two
models considered. Finally, we propagate our investigation to the inputs
required by N-body simulations in view of future applications, giving 
a table of the modifications required in both scenarios for taking into 
account the modified growth rate of perturbations and the strength of gravity. 
This work is organized as follows: we first devote
sec.(\ref{ws}) to Weyl scaling, recalling
how it is possible to relate EQ to CQ via a conformal transformation;
afterwards, in (sec.\ref{bkg}) we investigate
the background dynamics, comparing CQ in the simple case of a constant
coupling (\ref{bkg_cq}) and EQ in scalar tensor
theories (\ref{bkg_eq}); we then proceed by studying linear perturbations
(sec.\ref{lp}), again presented within
both cosmological approaches (\ref{lp_cq}, \ref{lp_eq}). We will then
point out in particular, how in the newtonian
limit (\ref{nl}) the Euler equation is modified (sec.\ref{nl_cq}, 
sec.\ref{nl_eq}), in view of N-body simulations to
be realized in both frameworks and for which we specify the required corrections (sec.\ref{nbody}). 
Finally in (sec.\ref{conclusions}) we compare the results and set our conclusions.

\section{Weyl scaling}
\label{ws}

Before handling a parallel analysis of the background and linear
perturbations in CQ and EQ models, it is compelling to
clarify the fact that the two frameworks are strictly related through a
conformal transformation called Weyl scaling
\cite{wands_1994, esposito-farese_polarski_2001, wetterich_1988,
maeda_1989, catena_etal_2007, doran_jaeckel_2002}.
It is important to stress that altering General Relativity via scalar
tensor theory (Jordan frame, JF)
is equivalent to coupling a scalar field universally with all matter
fields within General Relativity
(Einstein frame, EF). Although the two reference frames must lead to
identical observable effects,
the description of the same model can in fact rely on different and
unequally simple equations when seen in the two frames.
In this sense, simplicity and naturalness of one frame are not in general
preserved in the other one. Going from one
frame to the other and viceversa according to the model, profiting of this
frame equivalence, can therefore represent
a valuable tool to easily explore a wider set of choices for the coupling
functions and the potentials involved:
a nasty choice in one frame might, in other words, look much simpler in
the other frame, allowing us to easily stress
remarkably new features, more difficult to achieve in the frame of origin. 
A frame-independent approach has also been considered \cite{catena_etal_2007} leading to 
a reformulation of the Boltzmann equation and linear perturbation theory 
in terms of frame-independent quantities.
\\
Authors usually refer to CQ models when they consider a quintessence
scalar field coupled, in the Einstein frame,
to the dark matter fields. The action considered in this case usually
appears as follows:
\be
\label{CQ_action}
S = \int d^4x \sqrt{- g}\left[ \frac{1}{2\kappa} R - \frac{1}{2}
Z(\phi) \phi^{;\mu}\phi_{;\mu} - U(\phi) - m(\phi) \bar{\psi} \psi +
{\cal{L}}_{\rm{kin, \psi}}\right]\ ,
\ee
where $g$ is the determinant of the background metric, $R$ is the Ricci
scalar, $\kappa=8\pi G$ where $G$
represents the gravitational constant, as found in Cavendish like
experiments. In the assumption of flat
Friedmann Robertson Walker (FRW) cosmologies, the line element can be
written as $ ds^2 = -\, dt^2 + a^2(t)
\delta_{ij}dx^i dx^j \,, $ where $a(t)$ is the scale factor and $t$
represents the cosmic time variable.
The choice of $m(\phi)$ specifies the coupling to $\psi$ matter fields
while ${\cal{L}}_{\rm{kin, \psi}}$
includes kinetic contributions from all components different from $\phi$.
\\
In the Jordan frame, a scalar tensor theory in which EQ holds is in
general described by the following action:
\be
\label{EQ_action}  S = \int d^4x \sqrt{- g}\left[ \frac{1}{2 \kappa}
f(\phi,R) - \frac{1}{2}
Z(\phi)\phi^{;\mu}\phi_{;\mu} - U(\phi) - m_0 \bar{\psi} \psi +
{\cal{L}}_{\rm{kin, \psi}}\right]\ .
\ee
The scalar field $\phi$ characterizing the generalized gravitational
interaction has its kinetic energy and
potential specified by $Z(\phi)$ and $U(\phi)$, respectively.
In this case the mass $m_0$ is a constant and does not depend on the
scalar field $\phi$. Note that we consider
here natural units, $c=1$. Compared to general relativity, the Lagrangian
has been generalized by introducing an
explicit coupling between the Ricci scalar and the scalar field, achieved
by replacing the usual ricci scalar $R$
with the function $f(\phi,R)$. This new term, which has the effect of
introducing a spacetime
dependent gravitational constant, may either be interpreted as an explicit
coupling between the quintessence field
$\phi$ and gravity (or equivalently, in the Einstein frame, between dark
energy and matter), or as a pure geometrical
modification of general relativity admitting a non-linear dependence on
$R$ \cite{faraoni_2007}.
For simplicity, the classes of theories in which $f(\phi,R)$ assumes the
simple form $f(\phi,R)/2 = {\kappa} F(\phi)R/2$
have been widely considered in the context of dark energy cosmologies.
\\
Weyl scaling consists in a conformal transformation of the metric which,
joined to a redefinition of matter fields,
allows to rewrite action (\ref{EQ_action}) into (\ref{CQ_action}) or
viceversa:
\bea g_{\mu \nu} &=& A^2(\phi) \tilde{g}_{\mu \nu} \\
\nonumber \\
g^{\mu \nu} &=& A^{-2}(\phi) \tilde{g}^{\mu \nu}\ , \\
\nonumber \\
\sqrt{-g} &=&  A^4(\phi) \sqrt{-\tilde{g}}\ ,\\
\nonumber \\
R &=& [A(\phi)]^{-2} \{ \tilde{R} - 6 \tilde{g}^{\mu \nu}
(\ln{A})_{;\nu}(\ln{A})_{;\mu} \}\ , \\
\nonumber \\
\psi &=& A^{-3/2}(\phi) \tilde{\psi}\ ,
\eea
where we have used the $\tilde{}$ to identify quantities in the EF and
distinguish them from those in the JF.
Note also that the scaling factor $A(\phi)$ is related to the coupling
$F(\phi)$ via the following relation:
\be
\label{WS_A_def} A^2(\phi) = \frac{1}{\kappa F(\phi)}\ .
\ee
When applying Weyl scaling to the action (\ref{EQ_action}) with $Z(\phi) 1$ for simplicity, we obtain that
the rescaled action appears to be
\be
\label{WS_action}
\tilde{S} = \int d^4x \sqrt{- g}\left[ \frac{1}{2 \kappa}\tilde{R} -
\frac{1}{2}
\tilde{Z}(\phi) \phi^{;\mu}\phi_{;\mu} - \tilde{U}(\phi) - m(\phi)
\tilde{\bar{\psi}}{\tilde{ \psi}} +
\tilde{{\cal{L}}}_{\rm{kin, \tilde{\psi}}}\right]\ ,\ee
where
\be
\label{WS_tildeZ_def}
\tilde{Z}(\phi) = A^2(\phi) \left(1 + \frac{6}{\kappa} \frac{A_{,
\phi}^2}{A^4(\phi)} \right)\ ,
\ee
\be
\label{WS_tildeU_def}
\tilde{U}(\phi) = A^4(\phi) U(\phi)\ ,
\ee
and both the mass and the kinetic term for matter fields now also depend
on $\phi$:
\be
\label{WS_m_def} m(\phi) = m_0 A(\phi)
\ee
\be
\tilde{{\cal{L}}}_{\rm{kin, \tilde{\psi}}}(\tilde{\psi}, \phi,
\tilde{g}^{\mu \nu})  -i \tilde{\bar{\psi}} {\tilde {\gamma}}^\mu \nabla_\mu \tilde{\psi} +
i \frac{3}{2} \frac{A_{,\phi}}{A(\phi)} \tilde{\bar{\psi}} {\tilde
{\gamma}}^\mu \nabla_\mu \tilde{\psi}\ .
\ee
Note that the non standard kinetic term $\tilde{Z}$ in
(\ref{WS_tildeZ_def}) is always positive and one
could further re-absorb it by redefining the scalar field $\phi$
implicitly into $\varphi$ in such a way that:
\be
\label{varphi_def}
\left(\frac{\partial{\varphi}}{\partial \phi}\right)^2 \equiv
\tilde{Z}(\phi)\ .
\ee
This allows to rewrite action (\ref{WS_action}) in the usual form:
\be
\label{S_varphi}
\tilde{S} = \int d^4x \sqrt{- g}\left[ \frac{1}{2 \kappa}\tilde{R} -
\frac{1}{2}\varphi^{;\mu}\varphi_{;\mu} - \tilde{U}(\varphi) - m(\varphi)
\tilde{\bar{\psi}}{\tilde{ \psi}} +
\tilde{{\cal{L}}}_{\rm{kin, \tilde{\psi}}}\right]\ .
\ee
%In general, the function $\varphi(\phi)$ is not always described by an
%analytical expression. If however, we use the expression for
%$\tilde{Z}(\phi)$ coming out from the Weyl rescaling, we can analytically
%solve eq.(\ref{varphi_def}) thus obtaining:
%******
%TO CHECK!! (solo se si vuole passare da CQ a EQ):
%\\ \\
%\be \varphi = \frac{1}{2} \frac{A^2}{A_{, \phi}} {\cal D}(\phi) \left(1+
%\ln{A_{,\phi}}\right) -
%\frac{1}{2} \sqrt{\frac{6}{\kappa}} \ln{\left| \sqrt{\frac{6}{\kappa}}
%\frac{A_{, \phi}}{A^2} + {\cal D}(\phi) \right|}  \ee
%where
%\be {\cal D(\phi)} \equiv \sqrt{1 + \frac{6}{\kappa} \frac{A^2_{,
%\phi}}{A^4(\phi)}} \ee
%******
We have now all the means to find out, given a certain action
(\ref{EQ_action}) in the Jordan frame,
which is the equivalent action (\ref{S_varphi}) in the Einstein frame.
In the parallel analysis that will follow in this paper, we will
eventually refer to two particular
choices of the coupling in CQ and EQ scenarios. In particular, we will
consider two of the simplest
and most popular choices, that is to say a quadratic coupling in the
Jordan frame for EQ
\cite{zee_1979, wetterich_1988, perrotta_etal_2000, perrotta_etal_2004}
and an exponential coupling in the Einstein frame
for CQ \cite{amendola_2000, matarrese_etal_2003}. In order to have a feeling of a possible
comparison between the two different models,
we illustrate here how they look like in the same, Einstein, frame, after
Weyl scaling a quadratic EQ model. \\
In the case of EQ with a quadratic coupling, we will work with so called
`non-minimally coupled' theories, widely used in
previous works on this topic (see \cite{perrotta_etal_2000} and references
therein), in which $F(\phi)$ is the sum of a
dominant constant term plus a piece depending on $\phi$:
\be
\label{F_def_nmc}
F(\phi) \equiv \frac{ 1}{8 \pi G^{\ast}} + \xi \phi^2\ .
\ee
Here $\kappa_\ast \equiv 8 \pi G_\ast$, where $G_\ast$ represents the
`bare' gravitational constant
\cite{esposito-farese_polarski_2001}, which is in general different from
$G$ and is set in such a way
that locally $1/\kappa + \xi \phi^2_{local} = 1 / (8 \pi G)$ in order to
match local constraints on General
Relativity.%; indeed, in these theories the `strength' of gravity is
determined by a combination of $G_\ast$ and the local value of $f(\phi,
R)$ such that $G \equiv G_\ast {\frac{R} {f}} \left. \right|_{local}$.
In order to recover the equivalent action in the Einstein frame, we apply
equations
(\ref{WS_A_def}, \ref{WS_tildeZ_def}, \ref{WS_tildeU_def},
\ref{WS_m_def}), thus obtaining action (\ref{WS_action})
in which:
\bea
\tilde{Z}(\phi) &=& \frac{1}{\frac{G}{G^{\ast}} + \kappa \xi \phi^2} + 6
\kappa \xi^2 \phi^2\ ,
 \\
\nonumber \\ \tilde{U}(\phi) &=&  \frac{1}{\left[ \frac{G}{G^{\ast}} + \kappa \xi
\phi^2 \right]^2} U(\phi)\ ,
 \\
\nonumber \\ m(\phi) &=& \frac{m_0}{\sqrt{ \left|\frac{G}{G^{\ast}}+ \kappa \xi
\phi^2 \right|}}\ .
\eea
Notably, in this frame the coupling to matter fields, the kinetic term and
the potential do not look
as straightforward as in the original action. \\
Another very popular choice is represented by Induced Gravity (IG)
\cite{zee_1979, wetterich_1988, perrotta_etal_2000, holden_wands_2000}, in which only the
quadratic coupling
\be
F(\phi) = \xi \phi^2
\ee
is considered and no constant term is present. It can be easily seen, by
applying equations
(\ref{WS_A_def}, \ref{WS_tildeZ_def}, \ref{WS_tildeU_def}, \ref{WS_m_def})
that in this case
we recover action (\ref{WS_action}) in which:
\bea
\label{tildeZ_IG}
\tilde{Z}(\phi) &=& \frac{1}{\kappa \xi \phi^2} \left( 1 + 6 \xi \right)\ ,\\
\nonumber \\
 \label{tildeU_IG} \tilde{U}(\phi) &=&  \frac{1}{\kappa^2 \xi^2 \phi^4}
U(\phi)\ ,\\
\nonumber \\
\label{m_IG} m(\phi) &=& \frac{m_0}{\sqrt{\kappa \xi \phi^2}}\ .\eea
In this case it is also straightforward to solve equation
(\ref{varphi_def}) analytically,
obtaining
\be
\phi = D e^{C \varphi}\ ,
\ee
for $\phi >0$, where $C$ and $D$ are constants and we have defined $C
\equiv \sqrt{ \xi / (1 + 6 \xi)}$.
The latter expression, substituted in (\ref{tildeU_IG}, \ref{m_IG}) allows
to recover action (\ref{S_varphi})
in which:
\bea
\tilde{U}(\varphi) &=&  \frac{D^4}{\kappa^2 \xi^2}  \, e^{- C \varphi} \,
U(\phi(\varphi))\ ,\\
\nonumber \\
m(\varphi) &=& \frac{m_0 D}{\sqrt{\kappa \xi}} \, e^{-C \varphi}\ .\eea
Applying Weyl scaling to IG theories, leads, as expected
\cite{wetterich_1988, holden_wands_2000}, to CQ models
in which the mass of matter fields depends exponentially on the scalar
field $\varphi$.
We therefore recover the simplest and most popular choice in CQ
investigations \cite{amendola_2000}
and we will ourself adopt this choice in this paper when dealing with CQ
in the Einstein frame.
Furthermore, since $U(\phi)$ is usually almost constant in order to match
the present constraints
on the amount of Quintessence, the transformed potential in the Einstein
frame behaves as an exponential.
\\
As appears clear from the previous analysis, IG theories in the Jordan
frame (or equivalently CQ models in the
Einstein frame with an exponential coupling in the Lagrangian) have a
crucial difference from the non-minimal
coupling choice described by eq.(\ref{F_def_nmc}) in the Jordan frame.
Indeed, IG (or CQ with an exponential coupling) forces the coupling
constant $\xi$ to be positive in order to get
the right sign for attractive gravity in action (\ref{EQ_action}); on the
contrary, the choice (\ref{F_def_nmc})
also allows negative values of $\xi$ and can therefore lead to an opposite
correction to background and perturbation
dynamics, as we will see in detail in the following section.

% IG Weyl scaling
%we get, after Weyl scaling:
%
%\be \tilde{Z}(\phi) = \frac{1}{\kappa \xi \phi^2} \left( 1 + 6 \xi
%\right) \ee
%
%If $\frac {\left( 1 + 6 \xi \right)}{\xi} > 0$ then we can solve
%eq.(\ref{varphi_def}) for $\phi$ and rewrite the %action in terms of
%$\varphi$ with $\tilde{Z}(\varphi) = 1$:
%
%\be |\phi | = e^{C \frac{\varphi}{M}  + c_1} \ee
%where $c_1$ is a constant and we have defined
%\be C \equiv \sqrt{\left| {\frac{\xi}{1+6 \xi }} \right| } \ee
%\be m(\varphi) = \pm \frac{m_0}{\kappa \xi} e^{\left[{ C \varphi / M -
%c_1}\right]} \ee
%\be \tilde{U}(\varphi) = \frac{1}{\kappa^2 \xi^2} e^{\left[{-4 C \varphi
%/ M - 4 c_1}\right] } \, \, V(\phi(\varphi)) %\ee

\section{Background dynamics} 
\label{bkg}

In this Section we define the CQ and EQ models we study in the following, and review 
their dynamics for what concerns the cosmic expansion, highlighting the aspects which 
will be relevant for the behavior of linear perturbations. 

\subsection{Coupled dark energy}
\label{bkg_cq}

\begin{figure}[ht] 
\begin{minipage}{70mm}
\begin{center}
%\begin{picture}(185,235)(0,0)
\begin{picture}(185,180)(0,0)
%\put(90,-6){{$1+z$}}
\includegraphics[angle=0,width=7.5cm]{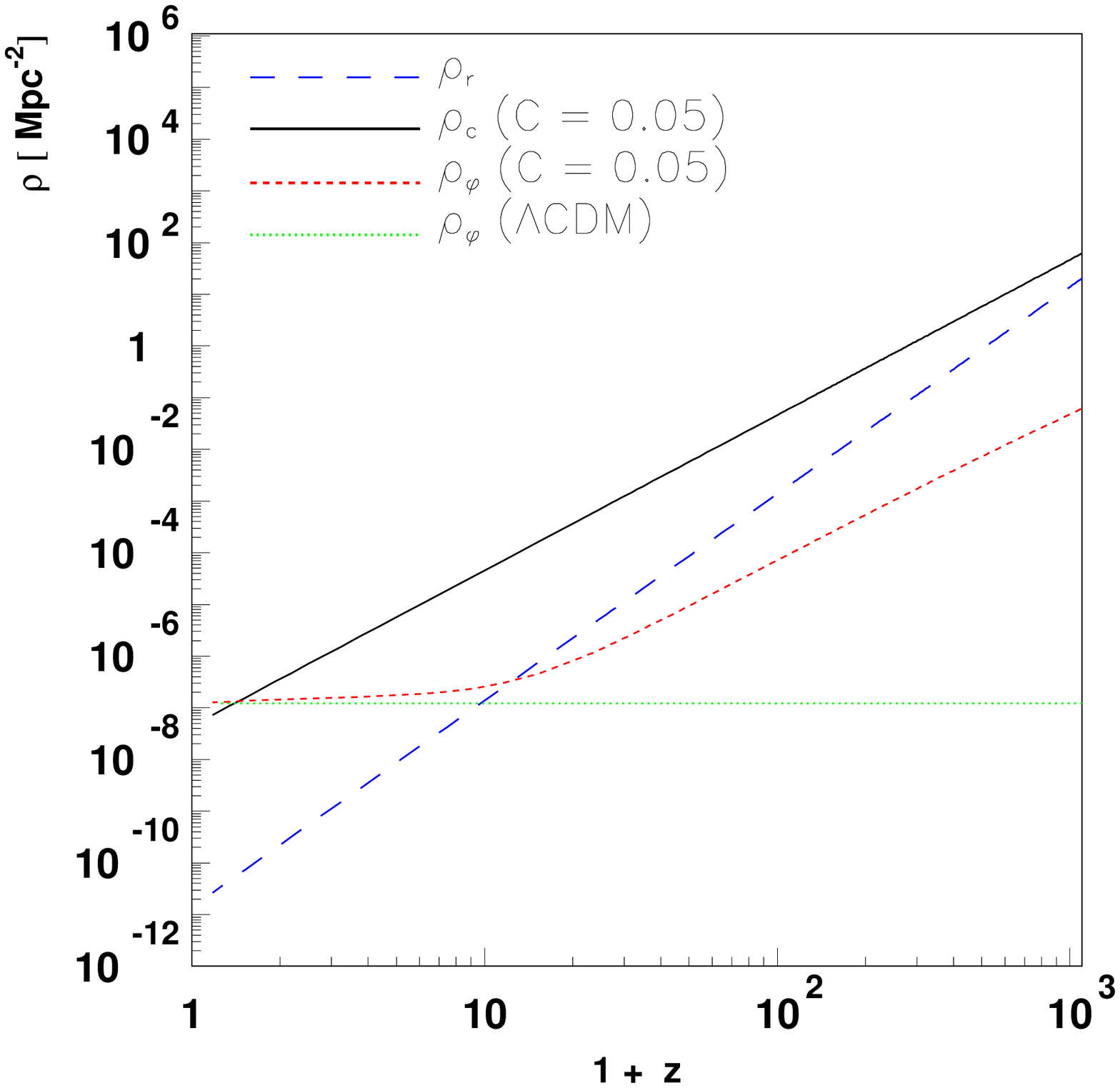}
\end{picture}
%\vspace{-4mm}
\label{fig1}
\end{center}
\end{minipage}
\begin{minipage}{70mm}
\begin{center}
\begin{picture}(185,180)(0,0)
%\put(90,-6){{$1+z$}}
\includegraphics[angle=0,width=7.5cm]{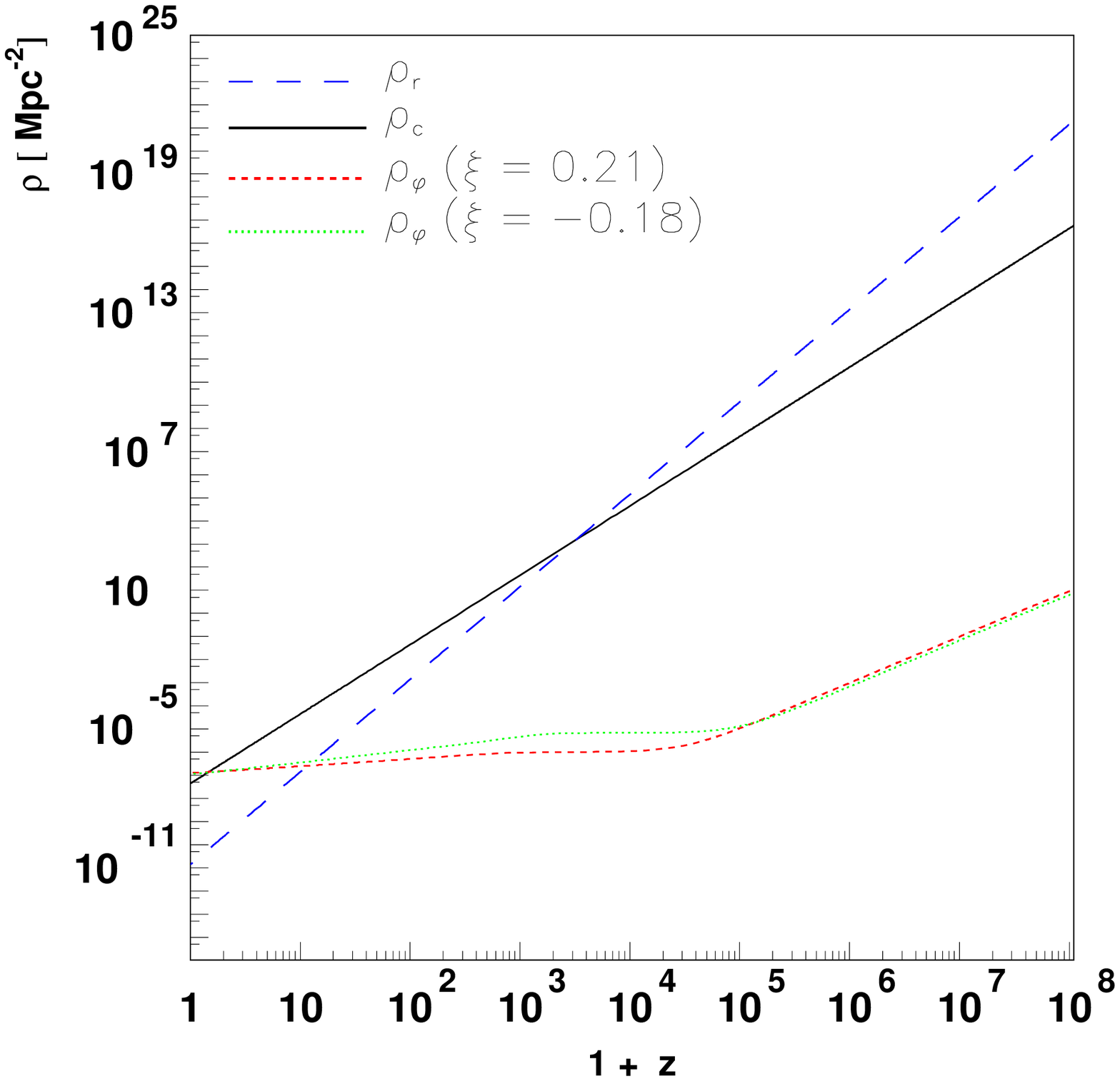}
\end{picture}
%\vspace{-4mm}
\label{fig2}
\end{center}
\end{minipage}
\vspace{4mm}
\caption{Energy densities of CDM 
and dark energy. The left panel shows the case of coupled dark energy for $C = 0.05$ 
as well as for $C = 0$ vs $1+z$, where $z$ is the redshift. The right panel shows the case of 
extended quintessence for both positive and negative coupling corresponding to $\omega_{JBD0} \sim 30$. 
During MDE the energy density of the scalar field (the non conserved one) is more enhanced in the case 
of coupled dark energy than in extended quintessence, in which the dynamics of the R-boost has its major 
influence during RDE. The plotted densities are in units of $Mpc^{-2}$.}\label{fig_rho}
\end{figure}

We will follow as much as possible previous works on this topic \cite{amendola_2000,amendola_2004}, 
considering the simplest models in which the dark energy scalar field and CDM are coupled 
via a constant coupling appearing in the conservation equations. \\
The equations used to describe the background evolution of each component $(\alpha)$ involved in the 
interaction follow from the consideration \cite{kodama_sasaki_1984} that the coupling can be treated 
as an external source acting on each stress energy tensor ${T{^\mu }}_{\nu (\alpha)}$ in such 
a way that the total stress energy tensor is conserved:
\bea 
\label{tensor_conserv_eq_c}
 T^{ \mu}_{ \nu ;  \mu \ (c)} &=& - C_c T_c \phi_{; \nu}
\\
\nonumber \\
\label{tensor_conserv_eq_phi} T^{ \mu}_{ \nu ;  \mu \ (\phi)} &=&  C_c T_c \phi_{; \nu}\ ,
\eea
where $C_c$ is a constant, $T_c$ is the trace of ${T{^\mu}}_{\nu (c)}$ and the subscripts $_{(c)}$, $_{(\phi)}$ 
stand for CDM and dark energy respectively. Baryons do not couple with dark matter or dark energy. 
The constant coupling term here used can be achieved starting from the following Lagrangian: 
\be  
S = \int d^4x \sqrt{- g}\left[ \frac{1}{2\kappa} R - \frac{1}{2}
\phi^{;\mu}\phi_{;\mu} - U(\phi) - m(\phi) \bar{\psi} \psi + {\cal{L}}_{\rm{kin, \psi}}\right]\ ,
\ee
in which the mass of cold dark matter fields depends exponentially on $\phi$ \cite{wetterich_1995}:
\be 
m(\phi) = m_0 e^{-C_c \phi}\ ,
\ee
thus corresponding, in the JF, to IG cosmologies, as seen at the end of the previous section.
Here, for simplicity, we do not include the baryonic component. 
\\
Equations (\ref{tensor_conserv_eq_c},\ref{tensor_conserv_eq_phi}) specialized for $\nu = 0$, 
provide the conservation equations for the energy densities of each component, given by 
${\rho}_c' = -3 {\cal{H}} {\rho}_c - C_c {\rho}_c {\phi}'$ and ${\rho}_{\phi}' = 
-3 {\cal{H}} {h}_{\phi} + C_c {\rho}_c {\phi}'$ for the two coupled species and 
${\rho}_{r}' = -4 {\cal{H}} {\rho}_{r}$ since $C_{r} = 0$ for radiation. 
%\bea
%\label{rho_conserv_eq_cdm} {\rho}_c' &=& -3 {\cal{H}} {\rho}_c - C_c {\rho}_c {\phi}'
%\\
%\label{rho_conserv_eq_phi} {\rho}_{\phi}' &=& -3 {\cal{H}} {h}_{\phi} + C_c {\rho}_c {\phi}'
%\\
%\label{rho_eq_rad} {\rho}_{r}' &=& -4 {\cal{H}} {\rho}_{r}
%\eea
$\rho_r$ takes into account the radiation contribution, including both photons and neutrinos, 
both decoupled from the other species. We have further defined 
${{h}}_\phi \equiv {{p}}_a + {{{\rho}}}_a = {{{{\phi}}^{'2}}} \slash {a^2}$
%\be \label{h_phi} {{h}}_\phi \equiv {{p}}_a + {{{\rho}}}_a = \frac{{{{\phi}}^{'2}}}{a^2} \ee
where the latter equality comes from the expression of the energy density $\rho_\phi$ defined 
from the $(0,0)$ component of the stress energy tensor present in (\ref{tensor_conserv_eq_phi}) 
and therefore given by the usual 
\be 
\label{rho_phi} 
{{{\rho}}}_{\phi} = \frac{1}{2}\frac{{{\phi}^ {'^2}}}{a^2} + U({\phi})\ ,
\ee 
where $U(\phi)$ is the potential in which the dark energy scalar field rolls. Note that $\rho_\phi$ 
is still equal to the standard one characterizing uncoupled quintessence models since, as we have seen, 
the coupling is provided in (\ref{tensor_conserv_eq_c}, \ref{tensor_conserv_eq_phi}) as an external 
contribution to the standard stress energy tensor; nevertheless, unlike what happens in the uncoupled 
quintessence case, only the sum $\rho_c + \rho_\phi$ is conserved while ${\rho_\phi}$ is not conserved 
by itself. \\
Analogously, the pressure of the quintessence scalar field is defined as:
\be \label{p_phi}
{{{p}}}_{\phi} = \frac{1}{2}\frac{{{\phi}^ {'^2}}}{a^2} - U({\phi})\ .
\ee
For what concerns the background dynamics, the evolution of the scalar field is described by the 
Klein Gordon equation, which reads as follows: 
\be  
\label{CQ_KG_eq}
{\phi}'' + 2 {\cal H} {\phi}' + a^2 U_{,\phi} = a^2 C_c{\rho}_c\ ,
\ee
where $U_{,\phi}$ is the derivative of the potential with respect to $\phi$. 
\\
The CDM conservation equation can be formally integrated out 
\be 
\label{rhoc_phi} 
\rho_c = \frac{\rho_{c0}}{a^3}e^{-C_c(\phi - \phi_0)}\, 
\ee
giving the dependence of $\rho_c$ from the scale parameter, once that the dynamics of $\phi(a)$ 
is known. \\
The evolution in time of the scale factor is provided by the usual Friedmann equation, so that:
\be 
\label{friedmann_eq} 
{\cal H}^2 \equiv  a^2 \, {\frac{\rho_c + \rho_\phi + \rho_r}{3}}\ . 
\ee
For convenience we have defined $\hat{\rho} \equiv 8 \pi G \rho = \rho/M^2$ and 
$\hat{\phi} \equiv \phi / M$ in units 
of the reduced Planck mass $M^2 \equiv 1/{8 \pi G} = {M_P}^2 / {8 \pi}$ where $G$ 
is the gravitational constant. The coupling constant $\hat{C} \equiv {C M}$ is therefore 
dimensionless; we have then omitted the $\hat{}$ and redefined $\hat{\rho} \equiv \rho$, 
$\hat{\phi} \equiv \phi$ and $\hat{C} \equiv C$ for simplicity. 
\\
A detailed analysis of the background has already been addressed in \cite{amendola_2000} in the case in
which $U(\phi)$ is an exponential and in \cite{maccio_etal_2004, amendola_2004, amendola_etal_2006}
for a more general treatment.
In particular, the critical points of the trajectories in the phase space of the system have been widely 
investigated in \cite{amendola_2000}. We just recall here that in the case in which both the potential 
$U(\phi)$ and radiation are neglected, 
%we can rewrite the Klein Gordon equation 
%(\ref{phi_KG_eq}) in terms of the convenient notation proposed in \cite{amendola_2000}: 
%\be \frac{dx}{d\alpha} = \frac{3}{2}(x^2 - 1)
%\left(x-\sqrt{\frac{2}{3}} C_c\right) \ee
%where we have changed the independent variable to $\alpha = \log{[a(\tau)]}$ and we have defined 
%\be \label{x_def} x = \frac{1}{\sqrt{6}} \frac{d \phi}{d \alpha} \ee
%From the latter expression 
it is straightforward to find the critical points of the system, corresponding to 
$d \phi_{crit} / d \alpha = (-\sqrt{6} , \, + \sqrt{6}, \, 2 \, C_c)$ where we have 
defined $\alpha = \log{[a(\tau)]}$. The third critical point is the only one providing a stable 
attractor during matter dominated era (MDE) \cite{amendola_2000} \footnote{The three critical points 
correspond to the last three critical points in table 1 in \cite{amendola_2000}.}
%. Along the attractor  \be x_{crit} = \sqrt{\frac{2}{3}} C_c \ee 
along which the evolution of the dark energy scalar field follows the
trajectory 
\be 
\label{phi_crit} 
\phi(a) = 2C_c \log{a} + \phi_0\ ,
\ee 
where we have put todays value of the scale parameter
equal to $a_0 = 1$. Substituting in (\ref{rhoc_phi}), we get:
\be 
\label{rhoc_a} 
\rho_c = \frac{\rho_{c0}}{a^3}a^{-2 {C_c}^2}\ . 
\ee
We can now make use of the Friedmann equation, specified in the case in which there is no radiation and the potential is negligible,
%\be \label{friedmann} {\cal H}_{MDE}^2 = a^2 \frac{\rho_c + \rho_\phi}{3} \ee
%in which $\rho_\phi$ is defined as in (\ref{rho_phi}),
to obtain the scale parameter dependence on the conformal time during MDE:  
\be 
\label{a_tau} 
a_{MDE}(\tau) \propto \tau^{\left[\frac{2}{1+2C_c^2}\right]} 
\ee as well as $\cal H(\tau)$:
\be 
\label{cq_H_tau}
{\cal H}_{MDE} (\tau) = \frac{2}{1+2C_c^2}\frac{1}{\tau}\ ,
\ee
or equivalently, in terms of the cosmic time 
$a_{MDE}(t) \propto t^{[{2} / {(3+2C_c^2)}]}$ and ${H}_{MDE} (t) = {2} / [({3+2C_c^2}){t}]$.
%\be \label{a_t} a_{MDE}(t) \propto t^{\frac{2}{3+2C_c^2}} \ee
%\be
%{H}_{MDE} (t) = \frac{2}{3+2C_c^2}\frac{1}{t}
%\ee
We stress here that, as we can see from the four latter expressions, the effect of the coupling 
on the Hubble parameter and on the scale factor during MDE, when the potential is negligible, 
does not depend on the sign of the constant $C_c$, both of them being only a function of $C_c^2$. 
With this choice of the coupling, we can thus only have a one direction effect, being that of 
slowing down the universe expansion with respect to the standard case in which $C_c = 0$. 
\\
We can numerically solve the equations describing the background, in order to obtain the full 
evolution of the energy densities of both cold dark matter and the scalar field, as shown in
fig.(\ref{fig_rho}). In the left panel, the energy densities of a cosmological constant (dotted), 
radiation (long dashed), CDM (solid) and coupled quintessence (dashed) in the case in which $C = 0.05$ 
have been plotted. We have used an inverse power law shape for the potential $U(\phi)$ which 
was first proposed in \cite{ratra_peebles_1988} and often used in quintessence models since 
then \cite{brax_martin_2000, zlatev_wang_etal_1999}: 
\be 
\label{U} 
U(\phi) = U_0\left(\frac{\phi_0}{\phi}\right)^\lambda\ ,
\ee 
where $\phi_0$ is the value of the scalar field today and $\lambda $
will be typically fixed to $ \sim 0.2$ in order to have a flat potential and a consequent 
equation of state for dark energy which is close to $- 1$ today, in order to be consistent 
with observations \cite{wmap_spergel_etal_2003, sdss_tegmark_etal_2004, hzss_riess_etal_2004}. 
Another popular choice is given by an exponential potential 
\cite{halliwell_1987, wetterich_1995, amendola_2000}, though we 
are not interested here in the explicit expression of $U(\phi)$, 
which does not modify our considerations. The amplitude of the potential (\ref{U}) and the initial 
amount of CDM are fixed in order to have their values today again in agreement with the present 
constraints \cite{wmap_spergel_etal_2003, sdss_tegmark_etal_2004, hzss_riess_etal_2004}.
As shown in fig.(\ref{fig_rho}) (left panel), the presence of the coupling has a drastic effect on the behavior 
of the scalar field energy density. In the uncoupled case quintessence behaves pretty much as a cosmological 
constant due to the rather flat potential. If however the coupling to CDM is active, $\phi$ accelerates and 
the dark energy density is attracted by the evolution of the species to which it is coupled: in particular, 
after a period of transition due to the particular choices of the initial conditions, 
${\bar{\rho}}_{\phi}$ is dragged by CDM to follow a trajectory which strictly mimics the CDM path, 
therefore behaving as an effective pressureless component. We will refer to this effect as `dark matter dragging' 
in analogy to the `gravitational dragging' first discussed in \cite{perrotta_baccigalupi_2002} in which 
the coupling of the scalar field to gravity via a modification of General Relativity imprinted a similar pattern to
the scalar field dynamics, both on background and on perturbations. In this scenario, the matter scaling is also 
modified by the coupling with the field; the resulting regime of cosmological expansion driven by the 
field and CDM tracking each other, is known as $\phi$CDM era, see \cite{amendola_2004} and references therein. 

In both models here considered, $\phi$ rolls down an effective potential provided by the coupling to dark matter 
(right hand side of eq.\ref{CQ_KG_eq}) or by modified gravity (right hand side of equation \ref{kg}, next paragraph). 
Note however that despite the analogy with gravitational dragging, during dark matter dragging 
the dark energy scalar field does not exactly slow roll on the effective potential, as it happens in the 
gravitational dragging case: when the coupling to dark matter is considered, $\phi$ acquires a strong acceleration
from the dark matter field, that makes terms like $\phi''$ give a non negligible contribution in equation 
(\ref{CQ_KG_eq}), an occurrence which is absent in the EQ case, as we now discuss. 

\subsection{Extended quintessence} 
\label{bkg_eq}

\begin{figure}[!hbtp]
\begin{center}
\begin{picture}(185,200)(0,0)
%\put(-10,180){{$H$}}
%\put(105,-6){{$z$}}
\epsfig{file=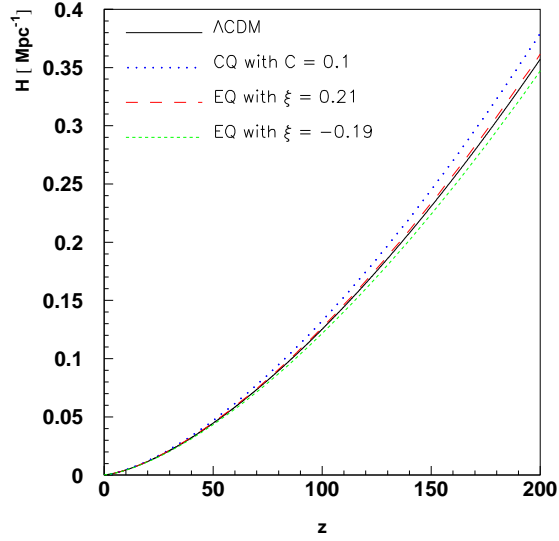,width=8.cm} 
\end{picture}
\caption{Hubble parameter vs redshift for CQ and EQ with a non minimal coupling (for positive and negative couplings). The case of $\Lambda$CDM is also shown for reference. The value of $\xi$ corresponds to $\omega_{JBD0} \sim 30$. $H$ is in units of $Mpc^{-1}$.}
\label{fig_hubble}
\end{center}
\end{figure}

As illustrated in the first section, scalar tensor theories are described by action (\ref{EQ_action}) 
that we rewrite here for convenience:
\be 
S = \int d^4x \sqrt{- g}\left[ \frac{1}{2\kappa} f(\phi,R) - \frac{1}{2} \phi^{;\mu}\phi_{;\mu} - 
U(\phi) - m_0 \bar{\psi} \psi + {\cal{L}}_{\rm{kin, \psi}}\right]\ . 
\ee
We have chosen $Z(\phi) = 1$ to simplify the notation. We further set $f(\phi,R) = {\kappa} F(\phi) R$ 
so that the function $F(\phi)$ fully specifies the explicit coupling to gravity via its dependence on some 
scalar field $\phi$. When the additional scalar field introduced in the coupling plays today the role of 
the dark energy field, leading to an accelerated expansion of the universe, we refer to it as extended 
quintessence, see \cite{matarrese_etal_2004} and references therein. The limit of general relativity is 
achieved with $\omega_{JBD} \gg 1$ where $\omega_{JBD}$ is defined as 
\be \omega_{JBD} \equiv
\frac{F}{{F_{\phi}}^2}\ ,
\ee 
and $F_{\phi}$ is the derivative of $F$ with respect to $\phi$. The current constraints set the present value 
$\omega_{JBD0}$ to be larger than 40000 in the Solar System \cite{bertotti_etal_2003}, and greater than 
about 120 from cosmology \cite{acquaviva_etal_2005}; being on markedly different spacetime scales, the 
two are considered as independent and complementary \cite{clifton_etal_2005}. 
Within these theories, the Friedmann equation describing the expansion of the Universe is modified with 
respect to the usual case: 
\be 
\label{frw} 
{\cal H}^2 = {\left(\frac{{a'}}{a}\right)}^2 = \frac{a^2}{3F}\left(\rho_{fluid} +
\frac{1}{2}\frac{{{\phi}'}^2}{a^2} + U(\phi) - \frac{3{\cal H}{{F '}}}{a^2}\right)\ . 
\ee 
Besides the extra term at the end of (\ref{frw}), which is usually quantitatively negligible, 
the most important effect to point out is the presence of the multiplying factor $1 / F$ substituting 
the usual $8 \pi G$; as a consequence, extended quintessence behave as theories in which gravity depends 
on a varying function and not on a gravitational constant anymore. 
Note that now all fluids but $\phi$ satisfy the usual conservation equation 
${\rho '}_\alpha = - 3 {\cal H} h_{\alpha}$ including CDM, which now scales as $\rho_c = {\rho_{c0}} / {a^3}$.
The evolution of the scalar field is explicited by the Klein Gordon equation, which again gets contribution 
from the coupling to gravity, enhancing the background dynamics of $\phi$ and giving rise to the R-boost effect 
illustrated in \cite{baccigalupi_etal_2000} 
\be 
\label{kg} 
{\phi ''} + 2{\cal H}{\phi '} + a^2 U_{, \phi} = \frac{a^2}{2}F_{, \phi}R  \ , 
\ee where 
$F_{, \phi}$ and $U_{, \phi}$ are the derivatives of the coupling $F(\phi)$ and of the quintessence potential $U(\phi)$ 
with respect to $\phi$. The term on the right hand side has the effect of altering the potential $U(\phi)$ into 
an effective potential in which the scalar field rolls. This effect, responsible for the gravitational dragging 
\cite{perrotta_baccigalupi_2002}, is analogue to the one produced by the right hand side of equation (\ref{CQ_KG_eq}) 
due to the coupling to matter fields, within the Einstein frame, as shown in fig.(\ref{fig_rho} right panel): 
here we plot the energy densities of radiation (long dashed), CDM (solid) and EQ in the case in which the coupling to gravity 
is positive (dashed) or negative (dotted). The energy density of the scalar field has a similar behavior during RDE 
(R-boost effect), independently of the sign of $\xi$, while the two patterns detach a bit during MDE. The plot has 
been obtained solving numerically the equations for the background in the case of non-minimal coupling, 
with $F(\phi)$ given by (\ref{F_def_nmc}). Again, the effect of the coupling on the background evolution is that of 
enhancing the amount of dark energy in the past, due to the gravitational dragging set on by the additional term 
in the Klein Gordon equation (\ref{kg}). Though in the quadratic case the gravitational dragging shows in particular 
during RDE, note that the time at which the dragging ends depends on the choice of the model and on the values of 
the coupling parameter. An exponential coupling in the lagrangian would have a bigger effect, lasting up to MDE and 
more recent times \cite{pettorino_etal_2004}. Although the phenomenology and the energy density scaling is clearly 
analogous for the CQ and EQ models plotted in fig.(\ref{fig_rho}), it is important to stress here that the sign of 
the coupling leads however to a different correction in the Hubble expansion parameter. Indeed, while in coupled dark 
energy the Hubble parameter does not depend on the sign of the coupling constant during MDE (eq.\ref{cq_H_tau}), in 
extended quintessence a positive and a negative coupling lead to different sign corrections, either enhancing or 
reducing the Hubble parameter with respect to the standard LCDM case. The effect is shown in fig.(\ref{fig_hubble}): 
here we plot the hubble parameter versus redshift for $LCDM$ (solid), coupled quintessence with $C = 0.1$ (dotted) and extended 
quintessence with a positive (long dashed) or negative (dashed) coupling. For the chosen coupled quintessence model, with a 
constant coupling, the hubble parameter is bigger in the past than the usual $LCDM$ case, independently of the sign 
of the coupling constant. The chosen extended quintessence case, instead, can lead to both higher and lower values 
in the past, depending on the sign of the coupling. Note also that the switch in sign does not lead to perfectly 
opposite contributions: when the coupling is negative the effect is bigger than in the case of a positive coupling 
with the same absolute value. A reason for this to happen might be that in the first case the extra term in the Klein 
Gordon equation adds to the usual potential, favoring an easier enhancement of the dynamics of the field; on the contrary, 
for a positive constant, the extra term contrasts the effect of $V(\phi)$, making it more difficult to enhance the dynamics 
of the field.

Finally, we recall that in this context, the Ricci scalar can be written in terms of the cosmological content of the
Universe \be 
\label{ricci} 
R=-\frac{1}{F}\left[-\rho_{fluid} + 3p_{fluid} + \frac{{\phi '}^2}{a^2} - 4 \, U +
3 \left(\frac{F''}{a^2}+ \frac{2 {\cal H} {F'}}{a^2} \right) \right]\vv 
\ee where $\rho_{fluid}$ and $p_{fluid}$ are the energy density and pressure summed
up over all possible cosmological components but $\phi$.

\section{Linear perturbations} 
\label{lp}

Here we perform a comparative analysis of the behavior of linear perturbations in the two scenarios. 
As we show in a moment, the main difference highlighted in the previous section persist at this 
level, as in the Newtonian limit which is outlined below. 

\subsection{Coupled dark energy} 
\label{lp_cq}

The linearized perturbation equations can be derived, following the notation of \cite{kodama_sasaki_1984}, 
from
\begin{equation}
\delta(T^{ \mu}_{ \nu ;  \mu \ (\alpha)}) = \delta(Q_{\nu \ (\alpha)})\ ,
\end{equation}
with $Q_{\nu \ (\alpha)} = (- a Q_{\alpha}, \bf{0})$ being the source term for the $\alpha-$component.
In particular, comparing with eq.(\ref{tensor_conserv_eq_c}) and eq.(\ref{tensor_conserv_eq_phi}),
for the case of a constant coupling here considered the source is defined as: 
\bea
Q_{\phi} &\equiv& \frac{C_c {\bar{\rho}}_c
\bar{\phi}'}{a} \\ \nonumber \\
Q_c &\equiv& - Q_{\phi} \pp\ 
\eea
For $\nu=0$ one gets the perturbed energy conservation equations, while for $\nu=i$ (spatial index) 
we can recover the first order Euler equations. In the Newtonian gauge in which the non-diagonal metric 
perturbations are fixed to zero, the linearized conservation equations for the matter component and for 
the scalar field, respectively, read as follows:
\bea
\label{delta_rho_c} \delta \rho_c' + 3 {\cal{H}} \delta \rho_c + \bar{\rho}_c k v_c +3\bar{\rho}_c {\bf{\Phi}}' &=&
- C_c(\bar{\rho}_c \delta {\phi}' + \bar{\phi}'
 \delta \rho_c )      \ \ \ ,
%\rm{Energy \ eq.\  for \  CDM} \ \ \ \ \ \ \ \ \ \ \ \ \ \ \ \ \ \ \ \ \ \
\\
\nonumber \\
\label{v_c} v_c'+({\cal{H}}- C_c \bar{\phi}')v_c &=& - k ({\bf{\Phi}} + C_c  \delta \phi)   \ \ \ ,
%\rm{Euler \  eq. \ for \ CDM}
%\ \ \ \ \ \ \ \ \ \ \ \ \ \ \ \ \ \ \ \ \ \
\\
\nonumber \\
\label{delta_rho_phi_eq} \delta \rho_{\phi}' + 3 {\cal{H}} (\delta \rho_{\phi}+ \delta p_{\phi}) +
k {\bar{h}}_\phi v_{\phi} + 3 {\bar{h}_\phi} {\bf{\Phi}}' &=&
C_c(\bar{\rho}_c \delta {\phi}' + \bar{\phi}' \delta \rho_c )  \ \ \ ,
%\rm{Energy \ eq. \
%  scalar \ field} \ \ \ \ \ \ \ \ \ \ \ \ \ \ \ \ \ \ \ \ \ \
\\
\nonumber \\
\label{v_phi}
\bar{h}_{\phi}v_{\phi}'+ (\bar{h}'_{\phi}+4{\cal{H}} \bar{h}_{\phi})v_{\phi} &=& k \delta p_{\phi}-
 k \bar{h}_{\phi}
{\bf{\Phi}} +C_c k \bar{\rho}_c \delta \phi \ \ \ ,
%\rm{Euler \ eq.  scalar \  field}
\\
\nonumber \\
\label{deltarho_r}
\delta \rho_r' + 4 {\cal{H}} \delta \rho_r + k \frac{4}{3} \bar{\rho}_r v_r +4 \bar{\rho}_r {\bf{\Phi}}' &=& 0 \ \ \ ,
\\
\nonumber \\
\label{v_r} \bar{\rho}_r v_r' - \frac{k}{4}{\delta \rho_r} &=& - k \bar{\rho}_r {\bf{\Phi}}   \ \ \ ,
%\rm{Euler \  eq. \ for \ rad}
%\ \ \ \ \ \ \ \ \ \ \ \ \ \ \ \ \ \ \ \ \ \
\eea
where k is the wavenumber, $v_\alpha$ is the velocity perturbation (along the direction of $\vec{k}$) for the $\alpha$ species and 
$\bf{\Phi}$ is the gravitational potential coming from the metric perturbation; in particular 
$\bf{\Phi} = -{\bf{\Psi}}$ if we neglect the anisotropic stress. Note that the latter is rigorously 
zero for the present scenario, as opposed to the EQ case as we will see in a moment. Thus 
the only source of that here is represented by neutrinos, due to their viscosity for having decoupled 
in a markedly relativistic regime. Also note that the two latter equations hold only if we neglect baryons, 
otherwise Thomson scattering with photons should be taken into account. A more convenient way to express 
equation (\ref{delta_rho_c}) is to rewrite it in terms of $\delta_c \equiv \delta \rho_c / \rho_c $ as 
\be 
\label{delta_c_eq} 
\delta_c' + kv_c + 3 {\bf{\Phi'}} = -C_c \delta \phi '\ .
\ee  
The gravitational potential $\bf{\Phi}$ is related to the background and perturbed densities 
and velocities of the various components via the following expression: 
\be \label{grav} {\bf{\Phi}} = \frac{a^2}{2 k^2} \rho \Delta =  \frac{a^2}{2 k^2} 
\left[ \delta {\rho}_c + \delta \rho_\phi + \delta \rho_r + 3
\frac{{\cal H}}{k} \left({\bar{\rho}}_c v_c + \frac{4}{3} \bar{\rho}_r v_r + {\bar{h}}_\phi v_\phi \right) \right] \ .
\ee 
Or, equivalently, in a more compact form and in terms of the $\Omega_{(\alpha)}$ as 
\be 
{\bf \Phi} = \frac{3}{2} \lambda^2 \left[ \sum_i{\Omega_i \left( \delta_i + 3 (1+\omega_i) \lambda v_i\right)} \right]\ ,
\ee
where we have defined $\lambda \equiv {\cal H}/k$ and where $\Omega_c = (a^2 \rho_c) / (3 {\cal H}^2)$ and the sum is 
extended to all components, including dark energy. Note that dark energy perturbations contribute to the gravitational 
potential and therefore on the dark matter perturbations through eq.(\ref{delta_c_eq}). \\
Deriving with respect to the conformal time, making use of equations (\ref{delta_rho_c}, \ref{v_c}, \ref{delta_rho_phi_eq}, 
\ref{v_phi}) as well as of the background conservation equations and substituting the expression of ${\bf{\Phi}}$ given 
by (\ref{grav}) we get the following simplified expression for ${\bf{\Phi '}}$: 
\be 
\label{dgrav_dtau} 
{\bf{\Phi '}} = -{\cal H} {\bf{\Phi}} - \frac{a^2}{2k}(\bar{\rho}_c v_c + \frac{4}{3} \bar{\rho}_r v_r + 
\bar{h}_{\phi} v_{\phi})\ .
\ee

The energy density perturbation for the dark energy scalar field can be expressed in 
terms of $\delta \phi$, $\phi$ and their derivatives in the following way: 
\be 
\label{delta_rho_phi} 
\delta\rho_\phi \equiv \rho_\phi \delta_\phi = \frac{1}{a^2}\left[\phi'(\delta \phi)' + 
a^2 U_{,\phi}\delta \phi + (\phi')^2 \bf{\Phi} \right] 
\ .
\ee
We recall that the latter expression, holding in the newtonian gauge, can be derived 
from ${{\delta T}^0_{0 (\phi)}}$ of the dark energy scalar field when comparing it 
to the usual one for a fluid of pressure $p_\phi$ and energy density $\rho_\phi$
(or, equivalently, directly perturbing (\ref{rho_phi})). 
Analogously, for the velocity perturbation in the newtonian gauge, one has 
\be 
\label{hphi_vphi} 
h_\phi v_\phi = \frac{k}{a^2} \phi' \delta \phi\ ,
\ee
which can be derived from the ${\delta T}^0_{j(\phi)}$ component of the perturbed stress 
energy tensor. Making use of the two latter expressions, $\bf{\Phi}$ and ${\bf{\Phi'}}$ can be 
rewritten as 
\be 
\label{grav_phi} {\bf{\Phi}} = \frac{ \left[ \phi' (\delta \phi)' +
3{\cal H}\phi' \delta \phi + a^2 U_{,\phi} \delta \phi  
 + 3  {\cal H}^2 \left[\Omega_c 
(\delta_c + 3 \lambda v_c)  +  \Omega_r 
(\delta_r + 4 \lambda v_r) \right] \right] } {2{k}^2 -  (\phi')^2}\ .
\ee

\be 
\label{dgrav_phi} 
{\bf{\Phi'}} = -\frac{1}{2}\left[\phi' \delta \phi + 2 {\cal H}{\bf{\Phi}} 
+ 3\frac{{\cal H}^2}{k} \left( \Omega_c v_c + \frac{4}{3} \Omega_r v_r \right) \right]\ .
\ee
Note also that due to the definition of the energy density and pressure of the quintessence
scalar field (\ref{rho_phi}, \ref{p_phi}) the perturbed pressure appearing in eq.(\ref{v_phi}) 
is simply equal to: 
\be \label{delta_p_phi} 
\delta p_\phi = \delta \rho_\phi - 2 U_\phi(\phi) \delta \phi\ .
\ee
As regard to the dark energy scalar field perturbation, $\delta \phi$ can be either written using 
(\ref{delta_rho_phi}) or it can be gained when solving the perturbed Klein Gordon equation: 
\be 
\label{delta_phi} 
\delta{\phi}'' + 2 {\cal H} \delta{\phi}' + (k^2 + a^2 U_{,\phi \phi}) \delta \phi + 4 \bar{\phi}'
{\bf{\Phi}} ' - 2 a^2 U_{,\phi} {\bf{\Phi}} = 3{\cal H}^2\Omega_c C_c 
\left[ \delta_c - 2 {\bf{\Phi}}\right] \ \ \ \ .
\ee

\begin{figure}[!hbtp]
\begin{center}
\begin{picture}(185,200)(0,0)
%\put(-10,180){{$\delta_c/a$}}
%\put(105,-6){{$1+z$}}
\epsfig{file=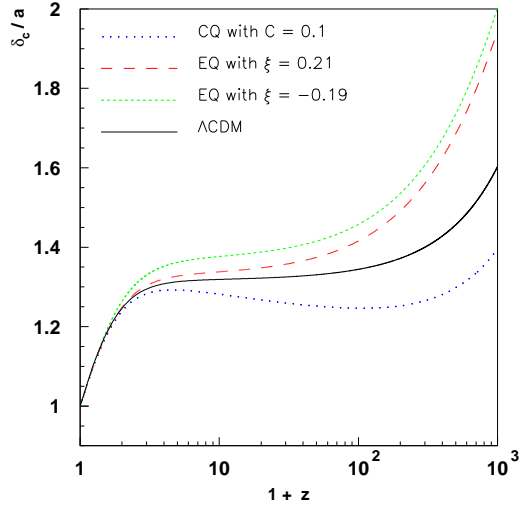,width=7.5truecm} 
\end{picture}
\caption{Growth factor as a function of redshift for coupled dark energy and extended quintessence with a 
non minimal coupling (for positive and negative couplings). The case of LCDM is also shown for reference. 
The value of $\xi$ corresponds to $\omega_{JBD0} \sim 30$.} 
\label{fig_growthfactor}
\end{center}
\end{figure}

In fig.(\ref{fig_growthfactor}) the density perturbation of CDM devided by the scale parameter and normalized 
to one today is shown; the latter quantity is the cosmological growth factor for CDM. The curves have been obtained 
by numerical integration and correspond to the case of CQ (dotted) as well as EQ case, for a positive (long dashed) or negative 
(dashed) coupling, as we will expose in detail in the following. The $\Lambda$CDM case is also shown for reference (solid): as it is well known, in the latter scenario the growth factor is 
almost constant during the Matter Dominated Era (MDE, $\delta_c \propto a$), while it increases going back in time due 
to the presence of radiation. The cosmological parameters of these scenarios are as defined in the previous section. 
We postpone further comments to this figure to the end of this section, when we expose the EQ case, too. 

\subsection{Extended quintessence} 
\label{lp_eq}

CDM variables follow the usual uncoupled equations, written here in newtonian gauge:
\be 
\label{EQ_deltac} 
\delta_c ' = -k v_c - 3 {\bf{\Phi '}}\ , \ee
\be \label{EQ_vc} v_c ' = -{\cal H}v_c +k{\bf{\Psi}}\ . 
\ee
Note that, in contrast with the CQ case, as we will see later, in EQ the two gravitational 
potentials $\bf{\Phi}$ and $\bf{\Psi}$ are not equal in module, since the anisotropic stress 
is in general different from zero. As we show now, however, this is not the major source of difference 
between the two scenarios. \\
As shown in \cite{faraoni_2000, perrotta_baccigalupi_2002} the perturbed Einstein equations can 
still be written formally as 
\be 
\label{EQ_einst_eq} 
\delta G^\mu_\nu = \kappa \delta T^\mu_\nu\ ,
\ee
where $G^\mu_\nu$ is the usual Einstein tensor and $\delta T^\mu_\nu$ is the total stress energy tensor, 
summing up all species: 
\be 
\delta T^\mu_\nu = \delta {T^\mu_\nu}_{(f)} + {\delta T^\mu_\nu}_{(\phi)}\ .
\ee
In particular  the conserved stress energy tensor for $\phi$ is given by 
\be 
{\delta T^\mu_\nu}_{(\phi)} = 
{\delta T^\mu_\nu}^{\, mc}{[\phi]} +{\delta T^\mu_\nu}^{\, nmc}{[\phi]} + {\delta T^\mu_\nu}^{\,grav}{[\phi]} \ ,\ee
where, in the newtonian gauge:
\be 
{\delta T^\mu_\nu}^{\, mc}{[\phi]} = 
\frac{1}{a^2}\left({\bf{\Psi}} {\phi '}^2 - \phi ' \delta \phi ' \right) -U_{,\phi}\delta \phi\, 
\ee
\be 
{\delta T^\mu_\nu}^{\, nmc}{[\phi]} = -\frac{3{\cal H}}{a^2} \left[2{\bf \Psi} - 
\frac{{\bf \Phi '}}{{\cal H}} \right] F' + \frac{3{\cal H}}{a^2}\delta F ' + 
\frac{F_{, \phi}R}{2}\delta \phi + \left(\frac{k^2}{a^2} - \frac{R}{2}\right) \delta F \ ,
\ee
\be 
\label{EQ_deltaTgrav} 
{\delta T^\mu_\nu}^{\, grav}{[\phi]} = \frac{3{\cal H}^2}{a^2}\delta F + \left(\frac{1} {\kappa}- F\right) \delta G^\mu_\nu \ .
\ee
Carrying the last term of expression (\ref{EQ_deltaTgrav}) to the left hand side in the perturbed Einstein equations, we can 
rewrite eqs(\ref{EQ_einst_eq}) as 
\be 
\label{EQ_einst_eq_F} 
F \delta G^\mu_\nu = {\delta T^\mu_\nu}^{\, mc}{[\phi]}  + {\delta T^\mu_\nu}^{\, nmc}{[\phi]} + 
\frac{3{\cal H}^2}{a^2}\delta F + {\delta T^\mu_\nu}_{(fluid)}\ .
\ee
The $(0,0)$ component of (\ref{EQ_einst_eq_F}) reads 
\bea 
\frac{2k^2}{a^2} {\bf \Phi} = 
\frac{1}{F}\left[\frac{6{\cal H}}{a^2} \left({\cal H} F + F' \right) {\bf \Psi} - 
\frac{3}{a^2}\left(2{\cal H}F + F' \right) {\bf \Phi '} + \right. \\ 
\left.  + \frac{1}{a^2} \left(\phi ' \delta \phi ' - {\phi '}^2 {\bf \Psi} + 
a^2 \left(U_{, \phi} -\frac{F_{,\phi}R}{2} \right) \delta \phi + \right. \right. \nonumber \\ 
\left. \left. - 3 {\cal H} \delta F' - \left(k^2 +3 {\cal H}^2 - a^2 \frac{R}{2}\right) 
\delta F \right) + \rho_f  \delta_f  \right]\ . \nonumber \eea

The $(0,j)$ component of (\ref{EQ_einst_eq_F}) gives the following equations:
\be 
\left[2 {\cal H} F + F' \right] {\bf \Psi} - 2 F {\bf \Phi '} = 
\phi ' \delta \phi + \delta F ' - {\cal H} \delta F + \frac{a^2}{k} {\delta T^0_j}_{fluid}\ .
\ee
Combining the two previous equations to get rid of ${\bf \Phi '}$, we get the Poisson equation 
in EQ cosmologies, which is 
\bea 
\label{EQ_poisson} 
\frac{2k^2}{a^2} {\bf \Phi} =  
\frac{1}{F} \left[ -\frac{1}{a^2} \left( {\phi '}^2 + \frac{3}{2} \frac{{F'}^2}{F} \right) {\bf \Psi} + \right. \\ 
\left. +  \frac{1}{a^2}  \left[ 3 \left({\cal H}  +  \frac{F'}{2F}\right) \phi ' \delta \phi + \phi ' \delta \phi ' + 
\frac{3}{2} \frac{F'}{F} \delta F' + \right. \right. \nonumber \\ 
\left. \left. - \left(k^2 + 6 {\cal H}^2 - \frac{a^2 R}{2} + \frac{3 {\cal H}}{2}  
\frac{F'}{F} \right) \delta F + a^2 \left( U_{, \phi}  - \frac{F_{, \phi}R}{2} \right) \delta \phi \right] + \right. \nonumber \\ 
\left. + \rho_f \delta_f + \frac{3}{k}\left({\cal H} + \frac{F'}{2F} \right) h_f v_f \right] \nonumber\ , \eea
where we have used that in the Newtonian gauge ${\delta T^0_j}_{fluid} = v_f h_f$. Finally, the Klein Gordon equation which 
describes the evolution of the EQ scalar field is 
\bea \label{EQ_deltaKG} \delta \phi '' + 2 {\cal H} \delta \phi' +
\left[ k^2 - \frac{a^2}{2}\left( F_{,\phi \phi}R - 2 U_{,
\phi \phi} \right) \right] \delta \phi = \\ = \phi' \left({\bf \Psi '} -3 {\bf \Phi '} \right) + \frac{a^2}{2} \left( F_{, \phi}R - 4 U_{,
\phi} \right) {\bf \Psi}  + \frac{a^2}{2} F_{, \phi} \delta R\ , \nonumber  \eea\\
where
\be 
\label{EQ_deltaR} 
\delta R = - \frac{2}{a^2} \left[ \left(3 {\cal
H} {\bf \Psi} - 3 {\bf \Phi '} \right)'  + 3{\cal H} \left(3 {\cal H}
{\bf \Psi} - 3 {\bf \Phi ' } \right)  - \left(k^2 - 3 {\cal H}' + 3
{\cal H}^2 \right) {\bf \Psi}  - 2 k^2 {\bf \Phi} \right] . \ee
These equations hold in general for any choice of the function $F(\phi)$. 
In order to solve the equations numerically, we again set the case of non-minimal coupling, 
in which (\ref{F_def_nmc}) is valid. \\

The results of the numerical integration are shown in fig.(\ref{fig_growthfactor}). As a reference, we normalize the growthfactor to a common value at present time. In the CQ case, the 
coupling to CDM has the effect of lowering the growthfactor with respect to the $\Lambda$CDM case; 
this means that for a fixed primordial normalization of the perturbations, the CQ structure formation 
is enhanced with respect to a $\Lambda$CDM case, independently on the sign of the coupling constant, 
as it is known from earlier works \cite{amendola_2004}. The EQ phenomenology is completely different, 
as the structure formation may be slower than the $\Lambda$CDM case, depending on the sign and magnitude of the
coupling constant. This may have interesting consequences for constraining these theories 
from the observations of cosmological structures in the mildly and full non-linear regimes, as we see 
in the next section. Note that, consistently with the analysis of the previous section, in the EQ 
case the departure from the $\Lambda$CDM case is again bigger for a negative coupling.

\section{Newtonian limit} 
\label{nl}

A coupling between dark energy and matter or gravity could indeed have observable effects on structure formation, 
due either to the interaction between species which behave differently with respect to gravity or because of a 
modification of gravity itself. We perform here a derivation of the Newtonian limit for cosmological perturbations 
in CQ and EQ scenarios \cite{amendola_2004, boisseau_etal_2000, schimd_etal_2005}. This part of the work represents the interface to numerical simulations of structure formation 
in these scenarios. Previous works on this subject considered the cases of CQ only \cite{maccio_etal_2004, amendola_2004}. 
We do expect to be able to track the differences outlined in the previous sections also in the present context. 

\subsection{Coupled dark energy} 
\label{nl_cq}

We strictly follow the notation of \cite{amendola_2004, maccio_etal_2004} and we specialize the previous general equations 
to the case in which $\lambda \equiv {\cal H}/k \ll 1$. This choice corresponds to the Newtonian limit, that is to say scales 
much smaller than the horizon. When the newtonian limit holds, the expressions of both the gravitational potential ${\bf{\Phi}}$ and 
its derivative ${\bf{\Phi '}}$ can be largely simplified and become approximately equal to: 
\be 
\label{grav_LLambda} 
{\bf{\Phi}} = \frac{1}{2{k}^2} \left[ \phi' (\delta \phi)' +
3{\cal H}\phi' \delta \phi + a^2 U_{,\phi} \delta \phi + 3 \Omega_c {\cal H}^2 \delta_c \right]\ ,
\ee
\be 
\label{dgrav_phi_LLambda} 
{\bf{\Phi'}} = -\frac{1}{2} \phi' \delta \phi - {\cal H}{\bf{\Phi}}\ . 
\ee
Substituting $\bf{\Phi'}$ in the perturbed Klein Gordon equation (\ref{delta_phi}), we get the following 
equation describing the evolution of $\phi$: 
\be 
\label{ddeltaphi_grav} \delta{\phi}'' + 2 {\cal H} \delta{\phi}' + (k^2 + 
a^2 U_{,\phi \phi} -2 \phi'^2) \delta \phi + {\bf{\Phi}}(- 4 {\cal H}{\phi}' - 2 a^2 U_{,\phi} 
+ 6{\cal H}^2\Omega_c C_c)  =  3{\cal H}^2 \Omega_c C_c \delta_c\ .
\ee
In the Newtonian limit $k \gg {\cal H}$ and the latter equation can be further simplified: the term 
containing ${\bf{\Phi}}$ can be neglected since ${\bf{\Phi}}$ is order $k^{-2}$; within the terms multiplying 
$\delta \phi$ , we neglect the contribution of the potential, which has effect only at very recent times, 
when the acceleration takes over; furthermore, the remaining piece  $k^2 -2 \phi'^2 $ can be rewritten as 
$ k^2 (1 -2 \lambda^2 (d\phi/d\alpha)^2) \sim k^2$ where $\alpha \equiv \log{a}$. Supposing that also 
$\delta \phi ''$ and $\delta \phi'$ can be neglected (this can be further checked a posteriori), 
eq.(\ref{ddeltaphi_grav}) in the newtonian limit reduces to: 
\be 
\label{KG_newtonian} 
\delta \phi \sim 3 \lambda^2 \Omega_c C_c \delta_c\ ,
\ee
stating that $\delta \phi$ is of order $\lambda^2$ and can therefore be neglected, together with its 
derivative, in eq.(\ref{grav_LLambda}) with respect to the last term:
\be 
\label{grav_poisson} 
{\bf{\Phi}} \sim \frac{3}{2} \lambda^2 \Omega_c \delta_c\ . 
\ee
By looking at the right hand side of eq.(\ref{v_c}) we can formally define the 
quantity:
\be 
\label{grav_c} {\bf{\Phi_c}} \equiv {\bf{\Phi}} + C_c \delta_\phi\ ,
\ee 
which reads
\be {\bf{\Phi_c}} = \frac{3}{2} \lambda^2 \Omega_c \delta_c(1+2C_c^2)\ ,
\ee 
where we have used the approximated expression (\ref{KG_newtonian}) for 
$\delta\phi$ obtained by solving the KG equation in the newtonian limit. 
The latter equation reads in real space as 
\be 
\nabla^2 {\bf{\Phi_c}} = - \frac{a^2}{2}\rho_c \delta_c (1+2C_c^2)\ ,
\ee
and if we explicit the units contained in $\rho$ \footnote{Recall that 
our energy densities are equal to $\hat{\rho} = 8\pi G \rho$ and we have redefined 
$\hat{\rho} = \rho$ for simplicity.}
we then have that $\bf{\Phi_c}$ behaves like a new gravitational potential satisfying:
\be 
\label{CQ_poisson} 
\nabla^2 {\bf{\Phi_c}} = - 4 \pi G_{\ast} a^2 {\rho_c} \delta_c\ ,
\ee
where we have defined $G_{\ast}$ as
\be 
\label{G_ast} 
G_\ast \equiv G (1+2 C_c^2)\ .
\ee
In this way, we formally recover the usual Poisson equation, where the effect of the 
coupling is partially included in the redefinition of the gravitational potential ${\bf{\Phi_c}}$ 
and partially in a varying gravitational constant $G_\ast$. 

The equations for the CDM component (\ref{delta_c_eq} and \ref{v_c}) can also be analogously 
simplified and read, in the Newtonian limit, as:
\be 
\label{delta_c_newtonian} 
{\delta_c}' = -k v_c\ ,
\ee
\be 
\label{v_c newtonian} 
v_c'+({\cal{H}}- C_c \bar{\phi}')v_c= - k {\bf{\Phi_c}}\ .
\ee
Deriving the first equation and combining it with the second, it is straightforward to obtain 
\be 
\delta_c'' + ( {\cal H} -C_c \phi') \delta_c' - \frac{3}{2}{\cal H}^2 \Omega_c \delta_c (1+2 C_c^2) = 0\ .
\ee 
Furthermore, equation (\ref{v_c newtonian}) allows us to clarify how the Euler equation, relevant 
for the interaction of dark matter particles within structure formation, changes due 
to the presence of a coupling between dark matter and dark energy (Einstein frame). 
Taking into account expression (\ref{grav_c}) 
and rewriting (\ref{v_c newtonian}) in real space coordinates, the equation finally reads 
as a modified Euler equation:
\be 
\nabla v_c ' + ({\cal H} - C_c \phi ') \nabla v_c + \frac{3}{2}{\cal H}^2
\Omega_c \delta_c (1 + 2 C_c^2) = 0\ .
\ee
Supposing now that CDM is concentrated in one particle of mass $m_c$ at a distance $r$ from a 
particle of mass $M_c$ at the origin we can rewrite the cold dark matter density contribution 
as 
\be 
\label{mass} 
\Omega_c \delta_c = \frac{8 \pi G M_c e^{-C_c(\phi-\phi_0)}{\delta}(0)}{3{\cal H}^2 a}\ ,
\ee
where we have used the fact that a non-relativistic particle at position $r$ has a density 
given by $m_c n {\delta(r)}$ (where ${\delta(r)}$ stands for the Dirac distribution) with 
number density $n = e^{-C_c (\phi - \phi_0)}$ given by equation (\ref{rhoc_phi}) and we have 
assumed that the density of the $M_c$ mass particle is much larger than $\rho_c$.
The Euler equation in cosmic time $dt = a d\tau$ then reads: 
\be 
\label{CQ_euler} 
\nabla \dot{v_c} = -\tilde{H} \nabla v_c - \frac{4 \pi \tilde{G} M_c \delta(0)}{a^2}\ ,
\ee
where we included the mass scaling in a new redefinition of the gravitational constant
\be 
\tilde{G} = G_\ast e^{-C_c(\phi - \phi_0)} =  G (1+2 C_c^2) e^{-C_c(\phi - \phi_0)}\ ,
\ee
and the further effect of the coupling on the friction term is a redefinition of the
Hubble constant:
\be 
\tilde{H} = H \left(1 - C_c \frac{\dot{\phi}}{H}\right)\ . 
\ee
Though the choice of including the exponential correction in a redefinition of the gravitational 
constant might be convenient for numerical purposes, we can as well rewrite eq.(\ref{CQ_euler}) as:
\be 
\label{CQ_euler_mass} 
\nabla \dot{v_c} = -\tilde{H} \nabla v_c - \frac{4 \pi {G_\ast} \tilde{M}_c \delta(0)}{a^2}\ ,
\ee
in which gravity is still governed by a gravitational (though modified) constant and the time 
dependent extra contribution is included in the mass $\tilde{M}_c \equiv M_c e^{-C_c(\phi - \phi_0)}$ 
of dark matter: when dark matter particles interact with each other, each dark matter particle sees 
the other one as possessing a `modified' mass which takes into account the non-standard scenario due 
to the coupling.\\ 
\begin{figure}[!hbtp]
\begin{center}
\begin{picture}(185,200)(0,0)
%\put(-25,160){{$\tilde{G}/G_*$}}
%\put(105,-4){{$z$}}
\epsfig{file=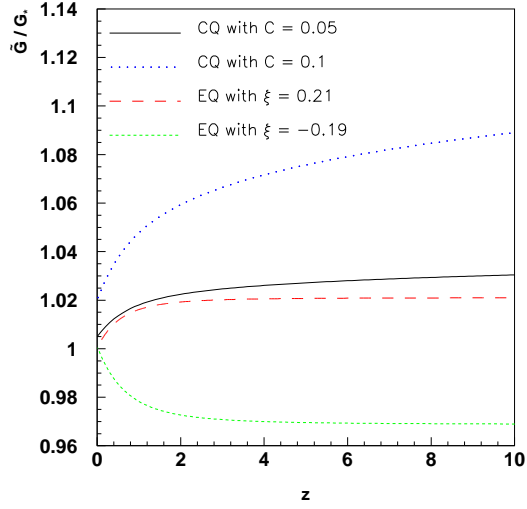,width=7.5truecm} 
\end{picture}
\caption{Gravitational correction as a function of redshift for CQ and EQ models. The value of $\xi$ corresponds to $\omega_{JBD_0} \sim 30$.} 
\label{fig4}
\end{center}
\end{figure}
Along the attractor (\ref{phi_crit}), the full correction to the gravitational constant and to the expansion 
assume the following expressions:
\be \label{gtildegforcq} \tilde{G} = G (1+2 C_c^2) a^{-2 C_c^2}\ , \ee
\be \tilde{H} = H \left(1 - 2 C_c^2 \right)\ , \ee
\be \tilde{M} = M_c a^{-2 C_c^2}\ . \ee
This clearly shows how the corrections behave regardless of the sign of the coupling constant. At the 
end of this Section, we'll focus on this aspect again, comparing the corrections to the strength of 
gravity in CQ and EQ.

\subsection{Extended quintessence} 
\label{nl_eq}

In the limit of small scales, in which ${k \gg {\cal H}}$ we can proceed in simplifying the equations 
for EQ in a way which is analogous to the one adopted in CQ. 
Neglecting time derivatives (negligible in the newtonian limit with respect to spatial derivatives) and 
the potential, the Poisson equation (\ref{EQ_poisson}) reduces to: 
\be 
\label{EQ_poisson_newtlim} 
\frac{2k^2}{a^2} F {\bf \Phi} = - \frac{k^2}{a^2} {F_{, \phi}} \delta \phi + \rho_f \delta_f\ ,
\ee
which reads, in real space, as:
\be 
\label{EQ_Poisson_Phi} 
F \nabla^2 {\bf \Phi} = - 4 \pi G a^2 \rho_f \delta_f - \frac{F_{, \phi}}{2}\nabla^2 \delta \phi\ . 
\ee
From the traceless part of the perturbed Einstein equations (component (i,j)) we get the gravitational 
potential relation 
\be 
\label{Phi_Psi_general} 
({\bf \Psi} + {\bf \Phi}) = - \frac{\delta F}{F}\ ,
\ee
so that in general the right hand side, representing the anisotropic stress, is different from zero in 
scalar tensor theories. 
Equation (\ref{EQ_Poisson_Phi}) can then be equivalently written, in terms of ${\bf \Psi}$, as 
\be 
\label{EQ_Poisson_Psi} 
F \nabla^2 {\bf \Psi} =  4 \pi G a^2 \rho_f \delta_f - \frac{F_{, \phi}}{2}\nabla^2 \delta \phi\ .
\ee
In the approximation we are considering, the perturbed Klein Gordon equation given by (\ref{EQ_deltaKG}) 
only receives contribution from 
\be 
\label{EQ_deltphi_newtlim}
\delta  \phi = {F_{, \phi}} \left[ {\bf \Psi} + 2 {\bf \Phi} \right]\ . 
\ee
Substituting (\ref{Phi_Psi_general}), we obtain: 
\be 
\delta \phi = \frac{F F_{, \phi}}{F + {F_{, \phi}}^2} {\bf \Phi} =- \frac{F F_{, \phi}}{ F + 2{F_{, \phi}}^2} {\bf \Psi}\ .
\ee
Further substituting in (\ref{EQ_poisson_newtlim}), the Poisson equation can be rewritten in the usual way
\be 
\label{EQ_Poisson_PhiE} 
\frac{2k^2}{a^2} {\bf \Phi_E} = {\frac{1}{F}} \rho_f \delta_f\ ,\ee
or, in real space, as:
\be 
\nabla^2 {\bf \Phi_E} = - \frac{4 \pi G}{F} a^2 \rho_f \delta_f\ ,
\ee
if we redefine the gravitational potential as 
\be 
\label{EQ_PhiE_def} 
{\bf \Phi_E} = \left(1 + \frac{1}{2}\frac{F_{, \phi}^2}{F + {F_{, \phi}}^2}\right){\bf \Phi}\ . 
\ee
Note that we could have equivalently substitute $\delta \phi ({\bf \Psi})$ in (\ref{EQ_Poisson_Psi}) 
thus obtaining 
\be  
\nabla^2 {\bf \Psi_E} =  \frac{4 \pi G}{F} a^2 \rho_f \delta_f\ ,
\ee
if we redefine the gravitational potential as: 
\be 
{\bf \Psi_E} = \left(1 - \frac{1}{2}\frac{F_{, \phi}^2}{F + 2{F_{, \phi}}^2}\right){\bf \Psi}\ .
\ee
The Euler equation (\ref{EQ_vc}) is then modified according to:
\be 
\label{EQ_vc_mod} 
v_c ' +{\cal H}v_c + k \left(1 + \frac{F_{, \phi}^2}{ F + {F_{, \phi}}^2}\right){\bf \Phi} = 0\ .
\ee
Substituting the expression for ${\bf \Phi}({\bf \Phi_E})$ given by (\ref{EQ_PhiE_def}) and making 
use of the Poisson equation (\ref{EQ_Poisson_PhiE}) we get 
\be 
\label{EQ_euler} 
v_c ' + {\cal H}  v_c + \frac{a^2}{2 k} \rho_f \delta_f \frac{2 ( F + 2 F_{, \phi}^2)}{F(2 F+3 F_{, \phi}^2)} = 0\ ,
\ee
which in real space reads as:
\be 
\label{EQ_vc_rhoc} 
\nabla {v '}_c  + {\cal  H} \nabla v_c + \frac{3}{2} {\cal H}^2 \Omega_f \delta_f \frac{2 (F + 2 F_{, \phi}^2)}{F (2 F+3F_{, \phi}^2)} = 0\ ,
\ee
where we have defined $\Omega_f \equiv \rho_f/\rho_{crit}$ and $\rho_{crit} \equiv 3 {\cal H}^2 F / a^2$. 
For comparison with the CQ case investigated in the previous paragraph, suppose now that the only other fluid present is CDM, concentrated in 
one particle of mass $m_c$ at a distance $r$ from a particle of mass $M_c$ at the origin, such that 
\be \Omega_c \delta_c = \frac{1}{F} \frac{M_c {\delta}(0)}{3{\cal H}^2 a}\ ,\ee
due to the fact that in this framework cold dark matter is uncoupled from dark energy and approximately scales in the usual way 
$\rho_c \propto a^{-3}$. Substituting the latter expression in (\ref{EQ_vc_rhoc}) we get 
\be 
\label{EQ_euler_cosmic} 
\nabla \dot{v}_c + H \nabla v_c + \frac{4 \pi \tilde{G} M_c \delta(0)}{a^2} = 0\ ,
\ee
in terms of the cosmic time, where we have included the extra contribution in the redefinition of the gravitational constant, 
which is now varying in time:
\be 
\label{EQ_Gtilde_def} 
\tilde{G} = \frac{2 ( F + 2F_{, \phi}^2)}{( 2 F+3F_{, \phi}^2)} \frac{1}{8 \pi F}\ .
\ee
The latter formalism is general for any choice of $F(\phi)$. For illustration we specify our final results to the non minimal 
coupling choice given by (\ref{F_def_nmc}). In this case 
\be 
\label{tildeG}
\tilde{G} =  \frac{ \left[ \frac{ 1}{8 \pi G_\ast}  + (1 + 8 \xi) \xi \phi^2 \right]}{ \left[\frac{1}{8 \pi G_\ast}  +  
(1  + 6 \xi) \xi \phi^2 \right]} \,  \frac{1}{\left[\frac{ 1}{G_\ast}  + 8 \pi \xi \phi^2 \right]}\ .
\ee
\\
For small values of the coupling, that is to say $\xi \ll 1$ the latter expression becomes:
\be \frac{\tilde{G}}{G_\ast} \sim 1 - 8 \pi G_\ast \xi \phi^2 \ee
which manifestely depends on the sign of the coupling $\xi$. 
In fig.(\ref{fig4}) we compare the behavior of the correction to the gravitational 
constant in CQ and EQ theories, for different values and sign of the coupling constants. 
As we stressed in this Section, in the CQ case the correction is independent on the 
sign of the coupling constant, just as expected since the theory corresponds to the case of 
induced gravity, which, as discussed in the first section, forces the sign dependence in order 
to mantain gravity as an attractive force; note also how also at present the value of the 
gravitational coupling is larger than the corresponding $\Lambda$CDM case, as it 
is evident from (\ref{gtildegforcq}). On the other hand, in the EQ case, 
the sign has the effect of increasing or decreasing the gravitational strength. 

\subsection{Recipe for Nbody users}
\label{nbody}
We conclude this section by summarizing the quantities needed as an input for N-body simulations 
when considering CQ models or EQ models as discussed in this paper. An N-body code which wants to 
take into account the modified gravitational strength between different species, as well as the modified 
perturbation growth rate, needs to read tables with the modification listed in table \ref{tab1} 
for both sets of cosmologies. The expansion history, represented by the Hubble expansion rate, 
needs also to be modified, although it is not listed in the table. \\
Three typologies of corrections are considered: the gravitational interaction between dark matter 
(DM) particles, the friction term due to a modified expansion and the mass of the interacting species. \\
When dealing with CQ models, all three corrections are active: the gravitational interaction between 
DM-DM particles changes, the friction term provides an extra contribution due to the coupling and the 
mass of the DM particles also changes with time and depending of the coupling. In the case in which we 
also consider baryons (B), non coupled to the scalar field $\phi$ or to CDM, the gravitational interaction 
between DM-B or B-B is the usual Newtonian one, since non coupled baryons satisfy the standard Euler equation. 
Furthermore, no correction to the baryon mass is required.\\
On the contrary, when EQ cosmologies are considered, only the gravitational correction needs to be applied. 
The effect of the coupling is all included in an effective gravitational `constant', which changes with time 
according to eq.({\ref{tildeG}}) for all the species considered. Both the friction term and the mass term remain unaltered.

\begin{table}
\begin{center}
%\newcommand{\mc}[3]{\multicolumn{#1}{#2}{#3}}
%\definecolor{tcA}{gray}{0.75}
%
\renewcommand{\arraystretch}{1.5}
\begin{tabular}{|ll | p {3 cm}| p{2 cm}| p{2 cm}|}\hline
{\bfseries       Correction type} &  & {\bfseries CQ} & {\bfseries EQ} \\ \hline
{\bfseries Gravity $\, ({\Delta G}/{G_\ast})$ } & {DM-DM} & {$ 1+2C_c^2$} & {${\tilde G}/{G_\ast} \,\, $} \\ \hline
 & {B-DM} & {$ 1$} & {$ {\tilde G}/{G_\ast} $} \\ \hline
× & {B-B} & {$ 1$} & {$ {\tilde G}/{G_\ast}$} \\ \hline
{\bfseries Friction $(\Delta {\cal H}/{\cal H} )$} & {\bfseries} × & {{$ 1 - \frac{C_c \phi'} {\cal H}$}} & {$1$} \\ \hline
{\bfseries Mass $ \,\,\,\,\,\,\,\, ({\Delta m}/{m})$ } & {DM} & {$ e^{-C_c(\phi - \phi_0)}$} & {$ 1 \,\,\,\,$} \\ \hline
× & {B} & {$ 1$} & {$1$} \\ \hline 
\end{tabular}
\end{center}
\caption{Summary of corrections required to run N-body simulations in CQ and EQ scenarios.}
\label{tab1}
\end{table}

\section{Conclusions}
\label{conclusions}

In this paper we performed a comparative analysis of the theoretical framework and 
phenomenology of cosmological perturbations in two of the main scenarios where the dark 
energy is explicitely coupled with dark matter (Coupled Quintessence, CQ) or gravity 
(Extended Quintessence, EQ). Weyl scaling the two theories, we fixed the expressions of the 
two scenarios in the Jordan and Einstein frames. This showed how they lead to markedly 
different phenomenologies, even if in the Jordan frame, they differ only by a constant term 
into the function multiplying the Ricci scalar: while the CQ case is bounded to produce effective 
corrections to gravity which increase its strength, the EQ case is able to provide corrections 
in both senses. This has important consequences which we highlight in the rest of the paper. \\
We show that indeed linear perturbation growth may actually accelerate or slow down in EQ 
cases with respect to a case in which the dark energy is uncoupled, while in CQ 
scenarios perturbations can be enhanced only. In order to set up a suitable framework 
for the realization of N-body simulations in these cosmologies, we derive the Newtonian 
limit for both, again highlighting the different behavior that the gravitational 
corrections possess, never neglecting the role of Quintessence fluctuations. 
Finally, we list the quantities which need to be modified in N-body codes for taking 
into account the corrections to the expansion, strength of gravity as perceived 
by different species, as well as the modified perturbation growth rate, in body 
scenarios, when performing simulations of non-linear structure formation. \\
Our work opens up the possibility of measuring the effect of non-minimal dark energy 
models in numerical simulations of structure formation, thus being able to constrain 
them through observations. N-body simulations nowadays reach the extension of 
thousands of Mpc, where indeed the interplay between large and small scale behavior of 
gravity really matters. Possible deviations from the ordinary behavior of gravity in 
General Relativity, which might be consistent with non-minimal dark energy models have 
to be detected coherently on all scales in order to provide convincing evidence. This 
capability can be built only by understanding properties and differences between the 
ways in which such deviations could manifest, developing capabilities for simulating 
and detecting them in large scale numerical simulations of structure formation and 
finally constraining them through observations. This work completes the first step in 
this path. 

\acknowledgments
We are grateful to F. Perrotta and J. Macher for helpful collaboration. We thank E.Linder, K.Dolag, L.Moscardini, M.Bartelmann, S.Matarrese, M.Meneghetti, C.Wetterich, G.Robbers for useful discussions. This work was supported by Progetto D4 Regione Friuli Venezia Giulia. VP is supported by the Alexander von Humboldt Foundation.

\end{document}